\documentclass[aps,prb,twocolumn,showpacs,amsmath,amssymb]{revtex4}
\usepackage{graphicx}
\usepackage{bm}
\def\be{\begin{equation}}       \def\ee{\end{equation}}
\def\bea{\begin{eqnarray}}      \def\eea{\end{eqnarray}}

\begin{document}

\title{Competing Order and Asymmetric Tunneling Spectrum in High Temperature Cuprate Superconductors}

\author{Jiangping Hu and Kangjun Seo}

\affiliation{Department of Physics, Purdue University, West Lafayette, Indiana 47907, USA}

\date{\today}

\begin{abstract}
We show that the asymmetric tunneling spectrum observed in the
Cuprate superconductors stems from the existence of a competing
order. The competition between the competing order and
superconductivity can create a charge depletion region near the
surface. The asymmetric response of the depletion region as the
function of the external voltage causes the asymmetric tunneling
spectrum. The effect is very general in a system which is near the
phase boundary of two competing states  favoring  different carrier
densities.  The asymmetry which has recently been observed in the
point-contact spectroscopy of the heavy fermion superconductor
CeCoIn$_5$  is another example of this effect.
\end{abstract}


\maketitle

\section{INTRODUCTION}

One important characteristic of the strongly correlated systems is
the possibility of the existence of many competing orders. To
identify these competing orders and study the interplay between them
are two major tasks in this field. In high temperature cuprate
superconductors,  competing orders, such as
magnetic,\cite{zhang,subir} stripe,\cite{kivelson,zaanen,carlson}
d-density wave, \cite{sudip} microscopic current orders, \cite{lee,
varma} and so on, have been proposed. Although it is still a matter
of debate, the competing orders have been widely considered to be
responsible for the unusual properties and the universal phase
diagrams of the materials. The competition between superconductivity
and antiferromagnetism or other orders  is also present in  other
strongly correlated electron systems such as heavy fermions.

In the cuprates, although the static global competing phases are
rarely detected and the direct transition between the
superconducting and the competing orders has not been observed,
recent STM experiments have suggested that a static order may
exist.\cite{hoffman,davis,arli} The competing phases are expected to
exist in the underdoped region. In the multiple layer cuprate
materials, the middle layer is naturally underdoped.\cite{DiStasio}
In this case, the competition between the competing orders and
superconductivity has been used to explain the behavior of Tc as a
function of the number of layers.\cite{chakravarty1} In this paper,
we show that the carrier distribution close to the surface can also
be naturally inhomogeneous along c-axis and the competing order is
generically enhanced at the surface. The competition between the
competing order and superconductivity can lead to an asymmetric
tunneling spectrum. The effect could be the natural explanation of
the asymmetry universally observed in the STM experiments in the
cuprates \cite{pan} and the metal contact tunneling experiments
\cite{park} in heavy fermion systems.

A particle-hole asymmetry is observed universally in different
cuprate compounds. In traditional BCS type superconductors,
tunneling spectrum is expected to be symmetric between negative and
positive bias voltage.  Such a large and universal asymmetry
observed in high temperature superconductors remains a challenging
puzzle to be solved.  A comprehensive theory to understand the
phenomena may be tied to the ultimate understanding of high
temperature superconductivity since there are only a few universal
features identified crossing different types of cuprate materials. A
few theoretical ideas based on the t-J model have been
proposed.\cite{wen,anderson,hirsch}

In this paper, we propose that the origin of this asymmetry is
indeed tied to one fundamental physics in cuprates, the existence of
the hidden competing order. We show that the asymmetry stems from
the competing order at surface, which has been detected by the same
STM measurement.\cite{hoffman,davis,arli} When an external voltage
is applied, the competing order near the surface can be weakened or
strengthened depending on the sign of voltage. Therefore, the region
of competing phases can be asymmetrically responded to the external
field. This effect leads to asymmetric tunneling matrix elements.
Such an effect is expected to be universal as soon as the competing
orders favor different doping concentrations from the
superconducting states. Interestingly, we notice that the similar
effect indeed has been observed in tunneling experiments in heavy
fermion systems.\cite{park} The sample used in the experiment is
also very close to the phase transition boundary between the AFM and
SC orders.

\begin{figure}
\includegraphics[width=7cm, height=5cm]{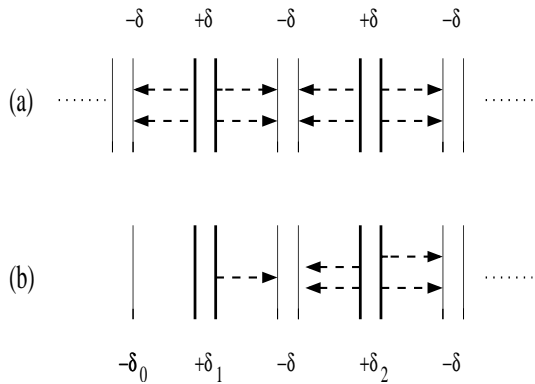}
\caption{\label{fig:layer} Schematic picture of the carrier
distribution and electrostatic field distribution along the c-axis
in a double layer structure. The solid and darker solid lines
represent the A-O and Cu-O planes respectively and the dash arrows
represent the electrostatic field. (a) and (b) represent the
distributions in the bulk and close to the surface respectively.}
\end{figure}

In a semiconductor, charges can be accumulated on the surface due to
the electronic surface states. In the bulk close to the surface, the
space charges can be induced to screen the surface charge. This
effect creates a space charge layer at the semiconductor interface.
In the cuprates, even without electronic surface states, a natural
charge depletion region near the surface can be created since the
carriers are introduced stoichiometrically. The number of the
carriers induced in the Cu-O layer close to the surface from the
neighboring layer A-O where A is La (Sr) in LSCO (BSCCO) can be very
different from that inside the bulk. On the surface, free charges
can be attached to the A-O plane to reduce the electrostatic energy
and effectively reduces the doping concentration. Such an effect
leads to the lower carrier density in the Cu-O plane close to the
surface. In Figure~\ref{fig:layer}, we plot a schematic picture to
show that  in the presence of the surface, the electrostatic  field
can be reduced close to the surface to save the electrostatic energy
which is the origin of the doping inhomogeneity in  the layered
compounds of the type Tl$_2$Ba$_2$Ca$_{n-1}$Cu$_n$O$_{2n+4}$ and
related structures.\cite{DiStasio} The effective carrier
contribution from the A-O layer on the surface is diminished  in
order to save the electrostatic energy. As a result,  the doping
concentration  on the Cu-O layer close to the surface is lower than
that in the bulk. Therefore, in the hole (electron) doped cuprates,
a hole (electron) depletion region can be created near the surface.
In the semiconductor, it is well known that an external voltage can
change the space charge distribution near the surface and the
depletion region has asymmetric response to an external electric
field. Thus one can expect that similar effect happens in the
cuprates, even though it is not clear how the asymmetric response
could create a large asymmetric STM tunneling spectrum if the sample
is simply in the superconducting state. In the presence of competing
orders, however, the effect could be crucial. The variation of the
carrier distribution could lead to a change of the configuration of
the competing orders near the surface. In particular, when the
competing orders are insulating, the tunneling amplitude can be
significantly modified by the change of the configuration of the
competing orders. Such an effect  could be accounted for the
observed asymmetric STM tunneling spectrum.

\section{MODEL AND FORMULATION}

We start to discuss about the physics in the structure shown in
Figure~\ref{fig:layer}. The total charge carrier energy near the
surface is given by $ F_c = \sum_{\langle i,j\rangle} V_{ij}\delta_i
\delta_j$, where $\langle i,j\rangle$ is a pair of any layers and
$V(x_i-x_j)$ are Coulomb energy between two layers if $i\neq j$ and
band energy if $i=j$. Since the physics is very general, we simply
assume  the sheet-charge model,\cite{DiStasio} in which the uniform
distribution of carriers in the Cu-O plane forms two-dimensional
sheets of charge. The charge in the A-O layers, except the one
exposed on the surface, are fixed and determined by doping
concentration. The charge in the layer at the surface can be freely
adjusted since it is exposed to outside. Therefore, we can consider
a uniformly doped superconductor with negative charge accumulated on
the surface with density $-\delta_0$. This surface charge induces a
non-uniform doping distribution of charges $\delta_i$ in the $i$th
Cu-O plane. Then the band energy under the carrier distribution,
$\rho_i = \delta_i / \delta $ along the $c$ axis can be written as

\be
    U_B = (\pi \hbar^2 / 2 m^\ast )\delta^2 \sum_i \rho_i^2,
\ee with $m^\ast \approx 4 m_e $.\cite{Krusin} And the electric
field is $ E(k) = ( 4 \pi /\epsilon ) \delta ( 1 -  \sum_j^k \rho_j
)$, where $\epsilon$ is the background dielectric constant. The
relevant electrostatic energy density is $U_e (i) = ( e^2 \epsilon /
8 \pi ) \sum_k^i  E_k ^2$. The total charge carrier energy  can be
written as

\begin{eqnarray}
F_c= \frac{2 \pi e^2 d_0 }{\epsilon} \delta^2 \left[
\frac{1}{C}\sum_{i} \rho_i^2 +  \sum_{i}  ( 1 -  \sum_k^i  \rho_k
)^2\right],
\end{eqnarray}
where $C = 4 m^\ast e^2 d_0 / \hbar^2 \epsilon$ and $d_0$ is the
distance between the adjacent planes.  Using the values $d_0 =
11.69$\AA~ and $\epsilon \approx 12\epsilon_0$, one obtains $C
\approx 7.4$.
\begin{figure}
\includegraphics[width=5cm,angle=-90]{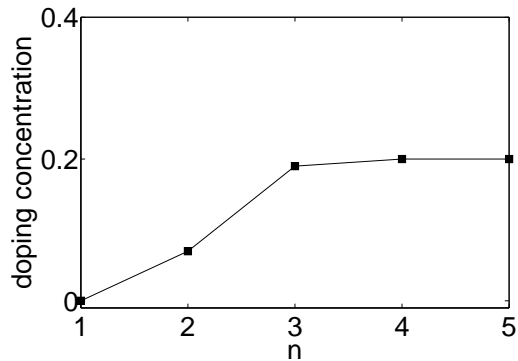}
\caption{\label{fig:doping} The doping concentration near the
surface. The average doping concentration inside the bulk is chosen
to be 0.2.}
\end{figure}

\begin{figure}
\includegraphics[width=8cm, height=10cm,angle=-90]{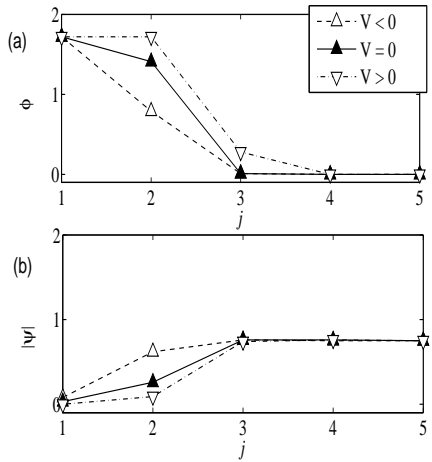}
\caption{\label{fig:ordervno0} The configurations of (a) the
competing and (b) superconducting order parameters when the voltage
is applied. V is the voltage dropped near the surface. The curves
are calculated at $V=100,0,-100$ meV respectively. The absolute
value of V is expected to be less than the bias voltage of the
sample with respect to the tip.}
\end{figure}

In the presence of a competing order, we adopt the Ginzburg-Landau
free energy of a multi-layer system given in Ref\cite{chakravarty1},

\bea\nonumber
F_{o}&=& \sum_j \left[ \alpha(\delta_j)|\psi_j|^2+\lambda |\psi_j|^4-\rho_c (\psi_j\psi^*_{j+1}+ c.c.) \right.\\
&&\left.+ \alpha'(\delta_j)\phi_j^2+\lambda' \phi_j^4 +
g|\psi|_j^2\phi_j^2 \right],\label{eq:freeenergy} \eea where
$\alpha(\delta_j )$ and $\alpha' (\delta_j)$ are functions of the
doping $\delta_j$ for each layer and all other parameters are
assumed to be constants. $\phi_i$ and $\psi_i$  are the competing
order and superconducting order parameters on the $i$th Cu-O plane
respectively. The total free energy of the system is given by $F_t=
F_c+F_o$. For a multi-layer system, in general, we have to resort
 to numerical method to minimize the total free energy.
 However, as an approximation, one can minimize  $F_c$  first with respect
to $\delta_i$ and obtain  the equilibrium charge distribution
$\{\delta_i\}$. After obtaining the charge distribution by
minimizing  $F_c$, we can minimize $F_o$ with respect to $\psi_j$
and $\phi_j$ to find the configurations of the competing and
superconducting order parameter as a function of layers. This simple
approach is proved to be reliable since Coulomb energy is usually
larger and only a few layer closest to surface are effected which we
will show later. In fact, we have performed an independent numerical
calculations to verify this by using of a variational Monte Carlo
calculation to search the global minimum of the total free energy.
The quantitative results for the distribution of charge density and
order parameter configuration are only slightly different from the
ones obtained through the above approximated approach.

 Without external voltage, it is
obvious that the A-O layer at the surface prefers zero charge
density, i.e. $\delta_0=0$, in order to minimize the charge energy.
Therefore, the first Cu-O near the surface will be underdoped in
general.  In the presence of external voltage, additional electric
field can be built near the surface, which leads to an effective
modification of the charge at the surface, i.e.,
$\delta(V)=\delta_0+ f(V)$, where $f(V)$ is zero if $V=0$ and
carries the same sign as the voltage $V$. Our natural assumption
here is that \bea
 f(V)= c_0 V.
\label{min} \eea If we assume that the entire voltage is dropped in
the competing order region which is roughly in one or two layers
along c-axis, the value of $c_0 $ can be estimated to be around
$\frac{\epsilon a^2}{d_0 e}$ which is $0.09(V^{-1})$ for the lattice
constant $a=4$\AA , $d_0=11.69$\AA, and $\epsilon = 12\epsilon_0$.
In the underdoped superconductor ($\delta_j < 0.2$, for all $j$), we
choose $\alpha(\delta_j) = 10(\delta_j - 0.3)$, $\alpha' (\delta_j)
= 27(\delta_j-0.22)$, $\lambda = \lambda' = 1$ and $g=1.2$. This
choice of the parameters gives the right shape and the magnitude of
the superconducting dome for the generic superconducting phase
diagram.\cite{chakravarty1}

\section{TUNNELING SPECTRUM}

The doping concentration at zero bias near the surface are
calculated and shown in Figure~\ref{fig:doping}, where the average
doping concentration in the bulk is taken to be $0.2$. This  figure
tells that only the first three layers closest to the surface are
effected. The doping concentration recovers very quickly to the
average values inside the bulk. The result also simplifies the
numerical calculation to search a global minimum for the total free
energy since only variables in  a few layers  need to be considered.

\begin{figure}
\includegraphics[width=8.5cm, height=6cm]{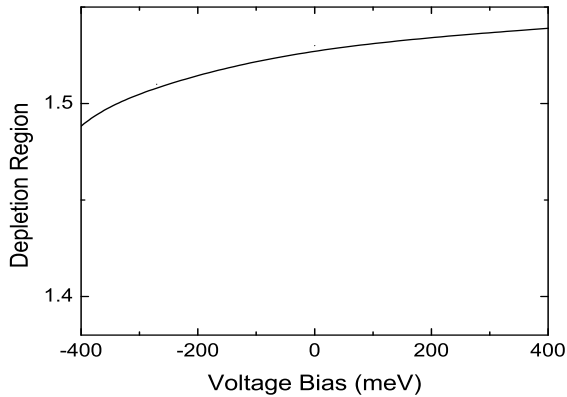}
\caption{\label{fig:depregion}The length of the charge depletion
region as the function of the voltage bias. The doping
concentration at bulk is set to be optimal.}
\end{figure}

 Figure~\ref{fig:ordervno0} shows how the configurations
of the order parameters respond to the applied voltage $V$, which is
the voltage dropped near the surface. The curves  in the figure are
calculated at $V=100,0,-100$ meV respectively. The absolute value of
$V$ is expected to be less than the bias voltage of the sample with
respect to the tip. The positive $V$ corresponds to electrons
tunneling into the sample. As the applied voltage increases from
negative to positive, the competing order on the second and third
layers increases whereas superconducting order  decreases on the
second layer. In the deep inside the sample, the superconducting
order  is uniform. Near the surface, however, the competing order
parameter is dominant over the superconductivity. This means that
the depletion region, where the competing order parameter $\phi$ is
dominant, increases with the increasing voltage from negative to
positive. The charge depletion region can be defined by the average
depth of the competing order parameter, $l_d$,
 \bea
 l_d= \sum_j j |\phi_j|^2 /\sum_j |\phi_j|^2.
 \label{eqn:dep_length}
\eea In Figure~\ref{fig:depregion}, we plot the length  of charge
depletion region, $l_d$, as the function of the voltage. The doping
concentration is chosen to be optimal.

 In tunneling
experiments, the tunneling matrix elements usually are assumed to
be constant. However, the presence of the charge depletion region
obviously leads to the variation of the matrix elements. The
increase of the charge depletion region near the surface with the
voltage bias $V$ implies the decrease of the conductance. Since
the matrix elements are determined by the overlapping of
wavefunctions between the tip and the measured material, the
change of the competing order near surface can result in a
significant variation of the tunneling matrix element. A slight
variation of competing order near surface could lead to large
change of  the tunneling-matrix element.

\begin{figure}
\includegraphics[width=5cm, height=8cm,angle=-90]{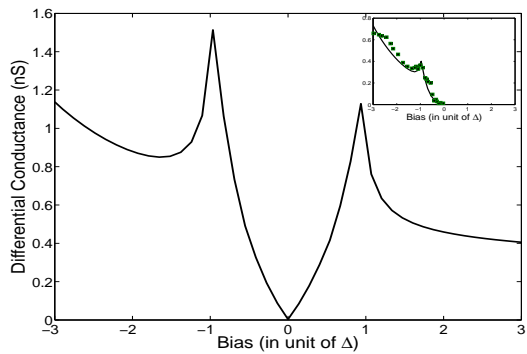}
\caption{\label{fig:depregion2} The tunneling spectrum calculated by
adding the effect of the depletion region in cuprates in
superconductring states.  The solid line in the inset shows the
difference between the spectrum in negative voltage and one in
positive voltage. The dashed dots are extracted experimental values
from Ref~\cite{pan}.}
\end{figure}

 The effect is universal as soon as a competing
order exists close to  superconducting phase, regardless of the
nature of the competing order. The effect is large if the competing
phase is insulating. One promising competing order is the
antiferromagnetic (AFM) order,\cite{zhang} which is an insulating
phase. It has been suggested that the transition between AFM and SC
orders are controlled by  chemical potential,\cite{zhang, project}
which can be changed by the doping concentration. Moreover, it has
also been argued that the superconducting materials are locked at
the AFM and SC transition point in the large range of doping
concentration,\cite{project} which inevitably results in a
microscopic phase separation. In this case, near the surface,  the
material should be very sensitive to
 external electric field since it effectively changes the chemical
potential. If the competing order is metallic, such as d-density
wave (DDW),\cite{sudip} the effect can be small. The value of
asymmetry is also expected to depend on the disorder which is
generically present in the cuprates. Although it is hard to make an
quantitative estimation since the different competing orders will
lead to different magnitude of the asymmetry, we can make a
reasonable assumption to obtain quantitative result to see whether
the experimental result can be naturally explained.

 In a simple
approximation, we can assume that the depletion region plays as a
tunnelling barrier in the STM measurement between the STM tip and
the superconducting phase. The energy of the barrier is determined
by the  single-particle energy gap $E_g$. Since the depletion region
is close to half filling,  $E_g$ should be expected in order of $U$
in the microscopic Hubbard model. This approximation leads to an
exponential dependence of the tunneling matrix element as a function
of the length of the depletion region,

\begin{eqnarray}
G(V) \approx G_0(V)e^{-\frac{2}{\hbar}\sqrt{2m^\ast E_g} l_d(V)d_0},
\end{eqnarray} where $G_0(V)$ is the tunneling conductance for a
d-BCS state which is assumed to be symmetric for negative and
positive bias voltages i.e. $G_0(V)=G_0(-V)$, $d_0$ is the lattice
constant along c-axis and $l_d(V)$ is the   depletion length in the
unit of lattice constant in c axis.  In Figure~\ref{fig:depregion2},
we calculated a typical spectrum following the above approximation.
The asymmetry between positive and negative bias voltage,
$G(-V)-G(V)$ is also shown in the inset. Amazingly, the experimental
data can be fitted to this crude approximation. The dotted points in
the inset of figure~\ref{fig:depregion2} are the extracted
experimental values from Ref.~\cite{pan} Using the result of
$l_d(V)$ calculated by Equation (\ref{eqn:dep_length}), there is
only one fitting parameter $E_g$. For the experimental
result,~\cite{pan} we obtain $E_g=0.79$ eV, which is indeed in the
order of $U$. Therefore, this result shows that the suppressed
superconducting order parameter by the dominance of competing order
parameter close to the surface of the sample can explain the
asymmetry of the tunneling conductance background of high
temperature superconductors.

\section{CONCLUSION AND DISCUSSION}

This theory provides several explicit predictions. First, there is
an onset temperature for
 the tunneling asymmetry since there is onset temperature for the
 competing order.  In cuprates, a natural onset temperature could
 be  $T^*$ associated with the pseudogap. We expect that the
 tunneling asymmetry disappears above $T^*$. Secondly, the
 tunneling asymmetry should be larger if the bias voltage is
 larger. It is a monotonically increasing function of the
  voltage, which is consistent with the experimental result of the tunneling
asymmetry as the function of voltage. The experimental result shows
that the asymmetry is small in low voltage and becomes larger and
larger as the voltage becomes large.\cite{pan} The theory predicts
that   there could be a large enhancement of the asymmetry at high
bias voltage if the nature of the transition between the competing
order and superconductivity is the first order transition. For
example, the transition between AFM and SC is typically first order.
Finally the asymmetry which has recently been observed in the
point-contact spectroscopy of the heavy fermion superconductor
CeCoIn$_5$ \cite{park} can also be naturally explained by this
effect since the material in this experiment is close to the
boundary of the AFM and SC phase transition.  Our theory predicts
that the asymmetry should be weakened when the material moves away
from the boundary toward superconducting zone by tuning the
pressure.

In conclusion, we have presented  that in the presence of
competing orders at surface, the tunneling matrix elements can
strongly depend on
 externally applied voltage.  This effect can lead to  a
natural explanation of the tunneling asymmetry observed in the STM
experiments. We have shown that this effect could exist
universally in  materials where the superconducting phase is close
to a competing order and our prediction  can be easily tested
experimentally.

The authors would like to thank  S. Chakravarty, S. Kivelson, S.
C. Zhang, E. W. Carlson, Congjun Wu and H.D. Chen
for valuable discussions. J. P. Hu would like to thank W. K. Park for
extremely useful discussion about his experiments.  This work is
supported by Purdue research funding.

\bibliography{asymmetry}

\end{document}